\documentclass[11pt]{article}
\usepackage[a4paper, footskip=15mm,footnotesep=9mm,nohead,dvips]{geometry}
\usepackage{amsmath}
\usepackage{amsfonts}
\usepackage{amssymb}
\usepackage{amsthm}
\usepackage{epsfig, lscape}

\renewcommand{\vec}{\boldsymbol}

\newtheorem{rem}{Remark}[section]   \newtheorem{exa}{Example}[section]   

\numberwithin{equation}{section}

\begin{document}

\title{The classical  Bertrand-Darboux problem}
 \author{Roman G. Smirnov\footnote{Electronic mail: smirnov@mathstat.dal.ca} \\
   Department of Mathematics and Statistics, 
Dalhousie University \\ Halifax, Nova Scotia,  Canada B3H 3J5 }
    \date{ }
 \maketitle

\begin{abstract}

The well-known  problem of classical mechanics considered by Bertrand (1857) and 
Darboux (1901) is reviewed in the context of Cartan's geometry. 
  
\end{abstract} 

UDK: 514.85; MSC: 37J35, 53C05


\bigskip 

\begin{array}[t]{cl}
 \hspace{1cm} \hfill\hfill\hfill\hfill \hspace{4cm} &   \mbox{\small\it ``... one thing seems certain. That is that the work of 
 }\\ 
 \hfill\hfill\hfill   &     \mbox{\small\it \'{E}lie  Cartan on connections, holonomy groups, and }\\
 \hfill\hfill\hfill & \mbox{\small\it homogeneous spaces is the source of all that is } \\
  \hfill\hfill &   \mbox{\small\it interesting in contemporary differential geometry.''}\\ 
 \hfill \hspace{1cm} \hfill \hspace{1cm} &  \mbox{K. Nomizu (cited in Guggenheimer \cite{Gug77}).}

\end{array}

\section{Introduction} \label{sec: intro}

In 1901 Darboux revisited \cite{Da01} the problem of mechanics  considered originally by Bertrand \cite{Be57} in 1857 (for a Russian translation of the 1901 Darboux article refer to \cite{BM04}). Later
on the results of this investigation due to Darboux were included almost verbatim in the classical text of Whittaker 
on analytical mechanics \cite{Wh37}. To this day Darboux' approach to the problem, which thus came to prominence with the classical  
Whittaker's book,  is widely used in the mathematical physics literature on integrable and superintegrable Hamiltonian systems 
(see, for example, \cite{Wo84, Wa00, KKM05}). Thus, the idea used by Darboux in \cite{Da01} to solve the Bertrand-Darboux problem was generalized to
study Hamiltonian systems defined in Euclidean spaces of higher dimensions (see \cite{Wa00} and the relevant references therein). In this view, one can assert that the classical 
Bertrand-Darboux problem is ``that special case that contains all germs of generality'' 
(in David Hilbert's words). Accordingly, the aim of this review article is to revisit the classical problem (which was only partially solved by Darboux, for a complete solution, using Darboux' approach, see \cite{AP83}) once again and present a solution utilising the fundamental ideas of  Cartan's geometry (see \cite{IL03} and the
references therein). The approach is based on both recent \cite{BMS01, MS02, JMP02, JMP04, SY04, DHMS, CMP05, JM05, HM06} and   classical \cite{E34, WF65} results that incorporate Cartan's geometry, in particular,  - the method of moving frames \cite{EC35,  FO98, FO99, PO99, PO05}, - into 
the study of classical Hamiltonian systems, including the Bertrand-Darboux problem that concerns classical Hamiltonian systems with two degrees of freedom defined in the Euclidean plane. 

More specifically, we target  to compare,  using modern  notations
and language, the two approaches  to the problem originally considered by 
Bertrand and Darboux, namely Darboux' approach and the approach via Cartan's geometry. This will enable us to conclude that Cartan's geometry
is the most general setting for solving the Bertrand-Darboux problem and its generalizations.

\section{A brief historical review of the Bertrand-Darboux problem} 

An adequate presentation of the Bertrand-Darboux problem should begin with a brief review of Liouville's paper of 1846 \cite{Li46}
to which  its origins can actually be traced. Recall that Liouville in \cite{Li46} studied canonical Hamilton's  equations governing
the motion of a particle on a curved surface defined by  a metric in isothermal coordinates under the influence of a potential energy depending on coordinates only. He observed that if the metric and the potential of the corresponding Hamiltonian function (total energy) admitted
special separable forms
\begin{equation}
H(u,v,p_u,p_v) = \frac{\frac{1}{2}(p_u^2 + p_v^2) + C(u) + D(v)}{A(u) + B(v)}
\label{H1}
\end{equation}
in some system of coordinates $(u,v)$, the Hamiltonian system defined by (\ref{H1}) could be solved by quadratures. Here  $A(u), B(v), 
C(u)$ and $D(v)$ are some arbitrary smooth functions.  Throughout
this paper the configuration space of the Hamiltonian systems in question is a two-dimensional Riemannian manifold $(M,\vec{g})$
(the pseudo-Riemannian case can be treated similarly \cite{MS02,JMP04, SY04}), and the phase space is the cotangent bundle $T^*M$. Note that the metric of
the kinetic part of (\ref{H1}) assumes the following covariant form 
\begin{equation}
\label{lm}
ds^2 = (A(u) + B(v))(du^2 + dv^2). 
\end{equation}
The form (\ref{H1}) of the Hamiltonian function also implies additive separation of variables for the associated Hamilton-Jacobi equation. The metric (\ref{lm}) and potential of the Hamiltonian (\ref{H1}) are said to be in the {\em Liouville forms}. The converse problem was considered in 1881 by 
Morera \cite{Mo81} who showed that if  a Hamiltonian system with two degrees of freedom defined by a natural Hamiltonian 
\begin{equation}
H(\vec{q},\vec{p}) = \frac{1}{2}g^{ij}p_ip_j + V(\vec{q}), \quad i,j = 1,2
\label{H2}
\end{equation}
could be solved within the framework of the Hamilton-Jacobi theory of separation of variables, then its metric $\vec{g}$ with 
the components $g^{ij},$ $i,j=1,2$ and the potential $V(\vec{q})$ assumed the Liouville forms with respect to special separable coordinates $(u,v)$. Here $\vec{q} = (q_1, q_2),$ $\vec{p} = (p_1, p_2)$ are the standard physical  position-momenta coordinates. We note that this conclusion does not exclude the possibility that the Hamiltonian system defined by (\ref{H2})
separates in some other system of coordinates with respect to which the metric $\vec{g}$ and the potential $V(\vec{q})$ are not
in the Liouville forms. Morera has also demonstrated that in the Euclidean plane $\mathbb{E}^2$ separation of variables for the Hamiltnonian (\ref{H1}) in  the Liouville form occurs in Cartesian, polar, parabolic and elliptic-hyperbolic coordinates. The equivalence established by Morera appears to be incomplete, since it is not clear where the separable coordinates $(u,v)$ are to be derived from. 
The answer to this question, at least partial, was given by Bertrand and Darboux \cite{Be57, Da01}. Recall that Bertrand considered Hamiltonian systems given by the natural Hamiltonian (\ref{H2}) and admitting an additional first integral satisfying certain conditions. In this view he considered the governing equations describing the motion of a particle in the Euclidean plane $\mathbb{E}^2$ under the influence of the  potential force defined by a function $V(\vec{q})$ dependent on the position coordinates $\vec{q} = (q^1, q^2)$. Thus, he assumed that
the Hamiltonian system defined by (\ref{H2}) admitted a first integral of motion of the form
\begin{equation}
\label{fi}
F(\vec{q}, \vec{p}) = K^{ij}p_ip_j + B^{k}p_{k} + U(\vec{q}), \quad i,j,k =1,2
\end{equation}
and then looked for the potential function $V(\vec{q})$ in (\ref{H2}) satisfying this condition. 
Bertrand considers the case of  $B^1 = B^2 = 0$ 
and then shows that the vanishing of the Poisson bracket $0 = \{H,F\}$ yields two further conditions, 
namely the {\em Killing tensor equation}
\begin{equation}
\label{KT}
[\vec{g}, \vec{K}] = 0, 
\end{equation}
and the {\em compatibility condition}
\begin{equation}
\label{CC}
{d}(\hat{\vec{K}}{d}V) = 0,
\end{equation}
where $[$ , $]$ denote the Schouten bracket \cite{Sc40}, the $(1,1)$-tensor $\hat{\vec{K}}$ is
 given by $\hat{\vec{K}} = \vec{K}\vec{g}^{-1}$.
\begin{rem}{\em In what follows we shall always assume $B^1 = B^2 = 0$ in (\ref{fi}).}
\end{rem}

The Killing tensor equation (\ref{KT}) puts in evidence that the functions $K^{ij} = K^{ij}(\vec{q})$ 
in (\ref{fi}) are the components of
a Killing tensor of valence two defined in the Euclidean plane $\mathbb{E}^2$. 
More specifically, solving the Killing tensor equation (\ref{KT}) with respect
to the standard Cartesian coordinates of the Eulidean metric 
\begin{equation}
\label{EM}
\vec{g} = \partial_1\odot \partial_1 + \partial_2\odot \partial_2
\end{equation}
 yields the following general form
\begin{equation}
\begin{array}{rcl}
\vec{K} & = & \displaystyle (\beta_1 + 2\beta_4q_2 + \beta_6q_2^2)\partial_1
\odot  \partial_1 \\ [0.3cm]
& & + \displaystyle (\beta_3 - \beta_4q_1 -
 \beta_5q_2 - \beta_6q_1q_2)\partial_1\odot \partial_2 \\ [0.3cm]
& & + \displaystyle (\beta_2 + 2\beta_5q_1+\beta_6q_1^2) \partial_2\odot
\partial_2,
\end{array}
\label{gKt}
\end{equation}
where $\odot$ denotes the symmetric tensor product, 
$\partial_1 = \frac{\partial}{\partial q_1}$, $\partial_2 = \frac{\partial}{\partial q_2}$  and the arbitrary constants $\beta_1,\ldots, \beta_6$ are
the constants of integration. In local coordinates the Killing tensor equation (\ref{KT}) is an overdetermined system of
PDEs and the formula (\ref{gKt}) demonstrates that the vector space of its solutions ${\cal K}^2(\mathbb{E}^2)$ 
is six-dimensional: 
$\dim\,{\cal K}^2(\mathbb{E}^2) = 6$. Bertrand next turns his attention to the compatibility condition (\ref{CC}). In terms of the
Cartesian coordinates $\vec{q} = (q_1,q_2)$ it assumes the following form: 
\begin{equation}
\label{BDE}
\begin{array}{l}
\displaystyle \left(\frac{\partial^2 V}{\partial q_1^2} - \frac{\partial^2 V}{\partial q_2^2}\right)(\beta_6q_1q_2 + \beta_4q_1 + \beta_5q_2  - \beta_3) + \\[0.3cm]
\displaystyle \frac{\partial^2 V}{\partial q_1\partial q_2}(\beta_6q^2_2 - \beta_6q^2_1 + 2\beta_4q_2 - 2\beta_5q_1 + \beta_1-\beta_2) + \\[0.3cm]
\displaystyle 3\frac{\partial V}{\partial q_1}(\beta_6 q_2 + \beta_4) + 3\frac{\partial V}{\partial q_2}(\beta_6q_1 + \beta_5) = 0.
\end{array}
\end{equation} 
The linear PDE of second order (\ref{BDE}) was later called the {\em Bertrand-Darboux PDE}. 
Solving (\ref{BDE}) for $V(\vec{q})$
amounts to finding admissible potentials of the Hamiltonian systems defined by (\ref{H2}) whose integrability is afforded
by the existence of first integrals (\ref{fi}) which are quadratic in the momenta. Bertrand restricted
his attention to the solutions having the form  
\begin{equation}
\label{BS}
V[(q_1 - u)^2 + (q_2 - v)^2], 
\end{equation}
thus recovering known results due to Euler and and Lagrange about the motion of a particle attracted by two fixed centres according 
to Newton's law. 

The main goal of the 1901 paper due to Darboux \cite{Da01} was to show that the linear PDE (\ref{BDE}) 
could be solved in ``full generality'', thus
yielding a general potential $V$ of the Hamiltonian (\ref{H2}) that was compatible with the first integral
 (\ref{fi}). Before solving the PDE (\ref{BDE}), Darboux ingeniously observes that it can be simplified without 
loss of generality. Indeed, 
by rotating and translating the axes one can simplify the general form of (\ref{gKt}),  thus bringing it to a certain 
{\em canonical} form. Note that this coordinate transformation has no effect on the dynamics of the system. 
In addition, Darboux excludes the case when $\beta_6 = 0$, taking $\beta_6 = 1/2$ and also assumes that the Killing tensor after rotating and translating
the axes assumes the following canonical form: 
\begin{equation}
\vec{K}  = \left(\beta_1  + \frac{1}{2}q_2^2\right)\partial_1
\odot  \partial_1
 - \frac{1}{2}q_1q_2\partial_1\odot
\partial_2 + \left(\beta_2 +\frac{1}{2}q_1^2\right) \partial_2\odot
\partial_2.
\label{cK}
\end{equation} 
Note, that the coordinates $q_1, q_2$ and the parameters $\beta_1, \beta_2$ in (\ref{cK}) 
 are not the same as the corresponding
coordinates and parameters in (\ref{gKt}). They are connected via the action of the isometry group
 (see (\ref{A1}) below) that consists of
rotations and translations of the underlying space $\mathbb{E}^2$. 
\begin{rem}{\rm The canonical form of the Killing tensor (\ref{cK}) can be simplified further. Indeed, one can introduce the following
Killing tensor, given by $\vec{\tilde{K}} = \vec{K} -\beta_2\vec{g}$: 
\begin{equation}
\vec{\tilde{K}} = \left(\beta_1-\beta_2  + \frac{1}{2}q_2^2\right)\partial_1
\odot  \partial_1
 - \frac{1}{2}q_1q_2\partial_1\odot \partial_2 +\frac{1}{2}q_1^2 \partial_2\odot \partial_2,
\label{ccK}
\end{equation}
where $\vec{K}$ is given by (\ref{cK}) and $\vec{g}$ denotes the contravariant metric of the Euclidean plane $\mathbb{E}^2$ given 
in terms of the Cartesian coordinates $\vec{q} = (q_1, q_2)$ (\ref{EM}).  Indeed, in terms of the  Hamiltonian $H$ (\ref{H2}) and the first integral
$F$ (\ref{fi}) determined by the canonical Killing tensor (\ref{cK}) this is equivalent to introducing the first integral $\tilde{F}$ which is quadratic in the momenta and given by $\tilde{F} = F -\beta_2H$. Now the difference $\beta_1-\beta_2$ can be treated as the only parameter entering the formula (\ref{ccK}). Moreover, without loss of generality, one can always assume $\beta_1 - \beta_2 > 0$.} 
\end{rem}
In what follows, Darboux proceeds to find the unknown $V$ by solving the PDE (\ref{CC}) determined by the canonical Killing tensor (\ref{cK}), using
the method of characteristics. Eventually he introduces special coordinates $(u,v)$ with respect to which
the PDE (\ref{CC}) is easily solved, yielding the general form of the potential $V$ in terms of $(u,v)$: 
\begin{equation}
V(u,v) = \frac{C(u) + D(v)}{u^2 - v^2}.
\label{V}
\end{equation}
One immediately recognizes that the potential $V$ (\ref{V}) is in the Liouville form  (\ref{H1}), so is the corresponding metric. Hence, the Hamiltonian
system can be solved via separation of variables in the corresponding Hamilton-Jacobi equation.
 Moreover, the separable coordinates $(u,v)$ are the parameters of confocal conics related to the 
original Cartesian coordinates $\vec{q} = (q_1,q_2)$ as follows: 
\begin{equation}
q_1 = k\cosh u\cos v, \quad q_2 = k\sinh u\sin v,
\label{eh}
\end{equation}
where $\beta_1 - \beta_2 = 2/k^2$, $\beta_1, \beta_2$ as in (\ref{ccK}). 
More specifically, the parameters $u$ and $v$ represent two families of confocal ellipses and
 hyperbolas intersecting each other orthogonally. In this view these orthogonal coordinates 
are called {\em elliptic-hyperbolic}. In view of the fact that they afford separation of variables 
in the corresponding Hamilton-Jacobi equation of the Hamiltonian system (\ref{H2}), we call them 
{\em orthogonally-separable coordinates}.  In conclusion to this brief review of the results due
 to Darboux, we note that he essentially solved two problems:
\begin{itemize}
 \item[(1)] First,  under certain assumptions  he employed an appropriate transformation
 that consists of rotations and translations of the space $\mathbb{E}^2$ to reduce  
a given Killing tensor in the form (\ref{gKt}) to  its canonical form  (\ref{cK}). Note this transformation
 of $\mathbb{E}^2$ iduces the corresponding transformation in the vector space of 
Killing two-tensors ${\cal K}^2(\mathbb{E}^2)$ (see (\ref{A2}) below).
 
 \item[(2)] Second, he diagonalized the resulting 
canonical Killing tensor,  thus producing the corresponding orthogonally-separable coordinates (\ref{eh}) generated by two families of confocal 
conics. 
\end{itemize}
\begin{rem}{\em The metric $\vec{g}$ and Killing two-tensor $\vec{K}$ that determine the Hamiltonian $H$ (\ref{H2}) and first integral 
$F$ (\ref{fi}) in terms of the orthogonally-separable coordinates $(u,v)$  are given by the following diagonal forms:
\begin{equation}
\vec{g} =  \frac{1}{k^2(\cosh^2 u - \cos^2v)}\left(\partial_u\odot\partial_u + \partial_v\odot\partial_v \right) 
\end{equation}
and 
\begin{equation}
\vec{K} = \frac{1}{k^2(\cosh^2u - \cos^2v)}\left(\cos^2v\partial_u\odot\partial_u +\cosh^2u \partial_v\odot \partial_v\right) 
\end{equation}
respectively, where $\partial_u = \frac{\partial}{\partial u}$, $\partial_v = \frac{\partial}{\partial v}$. 
}
\end{rem}
In what follows, we shall describe the above problems solved by Darboux from the viewpoint of Cartan's geometry, as well as consider
the cases that he excluded from his considerations. The following example (see  \cite{JMP02}, as well as the relevant references therein for more details) presents an illustration of how Darboux' method 
can be employed to find orthogonally-separable coordinates $(u,v)$ for a Hamiltonian system with two-degrees of freedom admitting
a first integral of the form (\ref{fi})
and thus solve it by quadratures as shown by Liouville \cite{Li46}. 
\begin{exa}[2nd Integrable Case of Yatsun] \label{exa1}
{\rm
 Consider a Hamiltonian system with two degrees of freedom defined in $\mathbb{E}^2$ by the following Hamiltonian:
 \begin{equation}
 \label{YP2}
 \begin{array}{rcl}
 H(\vec{q},\vec{p}) & = & \displaystyle \frac{1}{2} (p_1^2 + p_2^2)  -2 (q_1^4 + 2 q_1^2 q_2^2 + \frac{2\lambda}{g_2}q_2^4)
  \\[0.3cm] & &  + 4 (q_1^3 + q_1q_2^2) - 2(q_1^2 + q_2^2).
\end{array}
\end{equation}
It is known that the Hamiltonian system defined by (\ref{YP2}) is completely integrable if $g^2= 2\lambda$ admitting in this case the following additional
first integral independent of (\ref{YP2}), which is quadratic in the momenta:
\begin{equation}
 \label{f2}
 \begin{array}{rcl}
 \displaystyle F (\vec{q}, \vec{p}) & = & \displaystyle \left(q_2^2 + \frac{3}{4}\right)p_1^2 - (2q_1 -1)q_2p_1p_2 + 
(q_1-1)q_1p_2^2  -3q_1^4 - \\[0.3cm] & & 
2q_1^2q_2^2 + q_2^4 + 6q_1^3 + 2q_1q_2^2 - 3q_1^2.
 \end{array}
 \end{equation}
Observe that the Killing tensor $\vec{K}$ determined by (\ref{f2}) is given by 
\begin{equation}
\label{KY}
\begin{array}{rcl}
\vec{K} & = &  \displaystyle \left(\frac{3}{4} + q_2^2\right)\partial_1\odot \partial_1 + \left(\frac{1}{2}q_2 - q_1q_2\right)\partial_1 \odot \partial_2 \\[0.3cm] & & 
\displaystyle + \left(-q_1 + q_1^2\right) \partial_2\odot \partial_2.
\end{array}
\end{equation}
We compare the general formula (\ref{gKt}) with the formula (\ref{KY}), and conclude that in this case $\beta_1 = \frac{3}{4}$, 
$\beta_2 = \beta_3 = \beta_4 = 0$, $\beta_5 = - \frac{1}{2}$ and $\beta_6 = 1$. Thus, in this case the PDE (\ref{BDE}) implied by the compatibility condition (\ref{CC}) is given by the following formula:
\begin{equation}
\label{BDE1}
\begin{array}{l}
\displaystyle \left(\frac{\partial^2 V}{\partial q_1^2} - \frac{\partial^2 V}{\partial q_2^2}\right)(q_1q_2 - \frac{1}{2}q_2) + 
 \frac{\partial^2 V}{\partial q_1\partial q_2}(q^2_2 - q^2_1  + q_1 + \frac{3}{4}) + \\[0.5cm]
\displaystyle 3\frac{\partial V}{\partial q_1}q_2  + 3\frac{\partial V}{\partial q_2}(q_1 -\frac{1}{2}) = 0.
\end{array}
\end{equation}
We perform next the change of the variables $q_1 \rightarrow z + \frac{1}{2}$
 $q_2 \rightarrow y$, to obtain:
 \begin{equation}
 \label{V1}
  \left(\frac{\partial^2 V}{\partial z^2} - \frac{\partial ^2 V}{\partial y^2}\right) zy +
\frac{\partial ^2 V}{\partial z \partial y}(y^2 - z^2 + 1) + \frac{\partial V}{\partial
z}3y - \frac{\partial V}{\partial y}3z =0.
 \end{equation}
 Consider now the differential equation of the characteristics of $(\ref{V1})$:
 \begin{equation}
 zy(dy^2 - dz^2) + (z^2 - y^2 - 1) dzdy = 0.
 \label{ec}
\end{equation}
Introduce again the new variables $u := z^2$ and $v:=y^2$ to transform (\ref{ec}) into the following ODE:
\begin{equation}
\label{Cl}
\left(\frac{dv}{du}\right)^2u - v + \frac{dv}{du}(u - v - 1) = 0,
\end{equation}
which is a {\it Clairaut's equation}, having a general solution of the form
\begin{equation}
(m+1)(mz^2 - y^2) - m = 0
\label{m}
\end{equation}
in the original variables $z,y$. Re-write (\ref{m}) in terms of a new parameter $a$ to get
$$\frac{z^2}{a^2} + \frac{y^2}{a^2 - 1} = 1.$$
Observe that  the characteristic
curves of the PDE are two families of confocal conics. Therefore taking
the parameters of the confocal hyperbolas and ellipses as coordinates, we have
$$z= ab, \quad y =[(a^2 -1)(1-b^2)]^{1/2},$$
or
$$z = \cosh u \cos v , \quad y = \sinh u \sin v .$$
Our next step is to  write the PDE (\ref{V1}) in terms of new variables $a$
and $b$ to obtain:
$$(b^2 - a^2)\frac{\partial^2 V}{\partial a \partial b}
+2b\frac{\partial V}{\partial a} -2a\frac{\partial V}{\partial b} = 0,$$
which has a general  solution
$$V =\frac{C(a)  + D(b)}{a^2 - b^2}$$
in the Liouville form (\ref{H1}). Finally, we transform back to the original coordinates $q_1$ and $q_2$ to find
\begin{equation}
\left\{ \begin{array}{l}
q_1 = \frac{1}{2} + \cosh u  \cos v,\\
q_2 = \sinh u \sin v. \end{array}\right.
\label{ct}
 \end{equation}
Therefore the separable coordinates are of the {\em shifted  elliptic-hyperbolic type}. Using these new coordinates, one can solve
the corresponding Hamilton-Jacobi equation by separation of variables and ultimately solve the Hamiltonian 
system in question by quadratures (for more details see \cite{JMP02}).

}
\end{exa}

\section{Geometry of the Killing two-tensors} 
\label{GKT}

As is well-known, the geometry of the aforementioned four coordinate systems that 
afford separation of variables in the Hamilton-Jacobi equation
of a Hamiltonian system defined by (\ref{H2}) in $\mathbb{E}^2$ and admitting a first integral of motion 
(\ref{fi}) is induced by the geometry of the corresponding Killing two-tensor that
determines (\ref{fi}). More specifically, the four
coordinate systems, namely Cartesian, parabolic, polar and elliptic-hyperbolic, are generated by the eigenvectors (eigenvalues)
of the corresponding Killing two-tensor. Thus, in each case the two families of confocal conics are the integral 
curves of the eigenvectors. In this setting we exclude from our consideration the {\em trivial} Killing tensors, which
are the multiples of the metric $\vec{g}$ and as such do not generate orthogonal coordinate systems. Such Killing
tensors  determine the Hamiltonian (\ref{H2}) and its multiples. Each {\em non-trivial} Killing tensor has (almost 
everywhere) distinct and real eigenvalues. Another important feature (known already to Jacobi) 
of the four systems of coordinates is that
the Cartesian, parabolic and polar coordinate systems are degeneracies of the elliptic-hyperbolic coordinate system. 
Thus in the most general case a coordinate system generated by a non-trivial Killing tensor has two focii (i.e., the points
where the eigenvalues coincide). This is the case (considered by Darboux \cite{Da01}) of the elliptic-hyperbolic
coordinate system. When the distance between the focii is zero, one gets the polar coordinates. Next, the 
case when one of the focii
goes to infinity, corresponds to the parabolic coordinate system. Finally, when both focii are at infinity, the coordinate
system is Cartesian. Figure 1 depicts the four coordinate systems along with the corresponding coordinate transformations
to the Cartesian coordinates. 

\bigskip 

\begin{tabular}{cc}
 $\begin{array}{c}
 \includegraphics{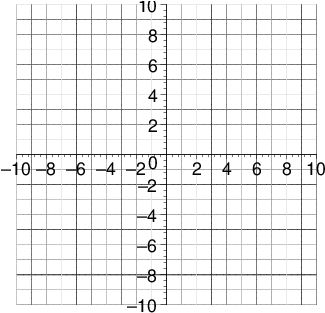} \\ 
 \mbox{I: Cartesian coordianes} \\
q_1 = u, \, q_2 = v
 \end{array}
 $
  & $\begin{array}{c}\includegraphics{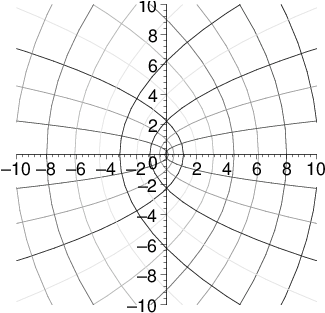}\\
  \mbox{II: Parabolic coordinates} \\
q_1 = 1/2(u^2 - v^2), \, q_2 = uv
  \end{array}
  $
  \end{tabular}

  \bigskip

  \begin{tabular}{cc} 
 $\begin{array}{c}
 \includegraphics{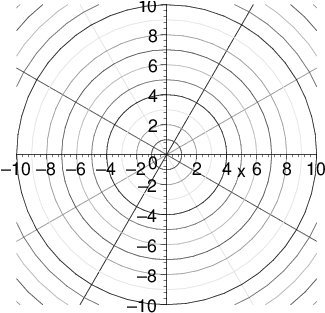}\\
 \mbox{III: Polar coordinates}\\
q_1 = u\cos v, \, q_2 = u\sin v
 \end{array}
 $
 & $ \begin{array}{c} 
 \includegraphics{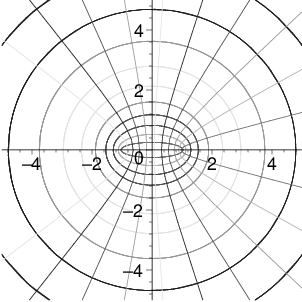} \\
 \mbox{ IV: Elliptic-hyperbolic coordinates}\\
q_1 = k\cosh u \cos v, \, q_2 = k \sinh u \sin v
 \end{array}
 $

 \end{tabular}
 
 \bigskip

\centerline{Figure 1: Families of confocal conics} 
\bigskip 

We conclude therefore that Darboux in his celebrated 1901 paper \cite{Da01} considered only the most general case, namely
orthogonal separation of variables in the case of the elliptic-hyporbolic coordinate system, excluding any degeneracies. 
When taking into account all of the cases, including the degeneracies, one is confronted with an {\em equivalence type problem}, that is for a 
given Hamiltonian system defined by (\ref{H2}) in $\mathbb{E}^2$, admitting a first integral quadratic in the momenta, the first
step is to determine what type of the four coordinate systems the corresponding Killing two-tensor generates. Once it is done, the
next step is to transform the Killing tensor and its orthogonal coordinate system to a canonical form. As we have seen
in Example \ref{YP2}, on account of the existence of a non-trivial potential $V$ in (\ref{H2}), 
the system of coordinates that afford separation of variables can be not in a {\em canonical} form.  Thus the elliptic-hyperbolic
system of coordinates that we have derived in Example \ref{YP2} is shifted (translated along the $q_1$-axis), that is affected by the action of
the corresponding isometry group. Mapping a given element of an equivalent class to the corresponding canonical form 
is done with the aid of the moving frames map. Therefore we conclude that the most natural framework  for solving the equivalence and canonical
forms problems is Cartan's geometry. This observation leads us to believe that the most general solution to the classical
Betrand-Darboux problem can be given in the framework of Cartan's geometry. This is the subject of considerations that follow.

\section{Cartan's geometry and Bertrand-Darboux' problem}
\label{CG} 
Let $I(\mathbb{E}^2)$ denote the Lie group of (orientation-preserving) isometries of  $\mathbb{E}^2$. It is 
a semi-direct product of $SO(2)$ (subgroup of orientation-preserving rotations) and $\mathbb{T}_2$ (subgroup of  translations). Recall, the action
$I(\mathbb{E}^2) \circlearrowright \mathbb{E}^2$ is transitive, yielding a description of $\mathbb{E}^2$ as a quotient
$I(\mathbb{E}^2)/SO_x(2)$, $x \in \mathbb{E}^2$. The action $I(\mathbb{E}^2) \circlearrowright {\cal K}^2(\mathbb{E}^2)$ 
(a general element of the six-dimentional vector space ${\cal K}^2(\mathbb{E}^2)$ is given by the generic formula (\ref{gKt})) is not transitive.
Solving the equivalence and canonical forms problem is this case is equivalent to analysing the orbits
of the vector space ${\cal K}^2(\mathbb{E}^2)$ under the action of $I(\mathbb{E}^2)$. The  discussion presented in 
Section \ref{GKT} suggests that there are four types of orbits generated by non-trivial Killing tensors of the vector
space ${\cal K}^2(\mathbb{E}^2)$. We shall exclude from further consideration the $0$-dimensional orbits
generated by trivial elements of ${\cal K}^2(\mathbb{E}^2)$ (since they do not generate coordinate systems and as such
have no relevance in Hamiltonian mechanics). 
The orbit  space ${\cal K}^2(\mathbb{E}^2)/I(\mathbb{E}^2)$ has a rather complicated structure, since 
 it contains $0$-, $1$-, $2$- and $3$-dimensional orbits (see below). It has a local differential structure
under the quotient topology. In addition,   the projection $\pi: \, {\cal K}^2(\mathbb{E}^2) \rightarrow {\cal K}^2(\mathbb{E}^2)/I(\mathbb{E}^2)$
 is a smooth fibration. 
The right action  $I(\mathbb{E}^2) \circlearrowright {\cal K}^2(\mathbb{E}^2)$ is (locally) free. 
Furthermore, for any point $x \in {\cal K}^2(\mathbb{E}^2)$ the corresponding orbit ${\cal O}_x \in
{\cal K}^2(\mathbb{E}^2)/I(\mathbb{E}^2)$ is an immersed submanifold in ${\cal K}^2(\mathbb{E}^2)$. Thus, 
$\xi = ({\cal K}^2(\mathbb{E}^2), \pi, {\cal K}^2(\mathbb{E}^2)/I(\mathbb{E}^2), I(\mathbb{E}^2))$ 
is a principle
$I(\mathbb{E}^2)$-bundle. For each orbit ${\cal O}_x$ passing through a non-trivial element 
$x \in {\cal K}^2(\mathbb{E}^2)$
we can sonsider the bundle of orthonormal frames of eigenvectors (eigenforms) of the elements of ${\cal O}_x$. 
The isometry group acts transitively on the bundle of orthonormal frames of ${\cal O}_x$. Thus, we
can employ the underlying ideas of Cartan's geometry, specifically - the {\em moving frames method} \cite{EC35, FO98, FO99, PO99, PO05}, - to solve the equivalence and canonical forms problem
for the action  $I(\mathbb{E}^2) \circlearrowright {\cal K}^2(\mathbb{E}^2)$.

Let $x \in {\cal K}^2(\mathbb{E}^2)$ be a non-trivial covariant Killing two-tensor $\vec{K}$. Recall \cite{BMS01}
 that in the {\em rigid
moving frame} of orthonormal eigenforms $E^1, E^2$ of $\vec{K}$ the components of  the metric $\vec{g}$ 
of $\mathbb{E}^2$ and
the Killing tensor $\vec{K}$ are given by the following formulas:
\begin{equation}
\label{mfe}
g_{ab} = \delta_{ab}E^a\odot E^b, \quad K_{ab} = \lambda_a\delta_{ab}E^a\odot E^b, \quad a,b = 1,2, 
\end{equation}
where $\delta_{ab}$ is the Kroneker delta, $\lambda_a,$ $a=1,2$ are the eigenvalues of $\vec{K}$. The dual 
vectors $E_1, E_2$ are the eigenvectors of $\vec{K}$. At a given point $x \in \mathbb{E}^2$ the two sets
 $\{E^1, E^2\}$ and $\{E_1,E_2\}$ form non-coordinate
bases of the cotangent and tangent spaces respectively. 
\begin{rem}{\em In the context of orbit analysis, chosing a rigid moving frame is equivalent to chosing
an appropriate cross-section (see below). Hence, this technique will yield canonical forms of the orbits of the
action $I(\mathbb{E}^2) \circlearrowright {\cal K}^2(\mathbb{E}^2)$.}
\end{rem}

Now we proceed to introduce the basic equations of Cartan's geometry. 
The equations for structure functions $C^c{}_{ab}, a, b, c = 1,2$ are given by 
\begin{equation}
[E_a,E_b] = C^c{}_{ab}E_c \quad \mbox{or} \quad dE^a = \frac{1}{2}C^{a}{}_{bc}E^b\wedge E^c.
\label{C}
\end{equation}
We intorduce  the connection coefficients $\Gamma$ as follows: 
\begin{equation}
\nabla_{E_a}E_b = \Gamma_{ab}{}^cE_c, \quad \nabla_{E_c}E^b = - \Gamma_{cd}{}^bE^d,
\label{Gamma}
\end{equation}
where $\nabla$ denotes the Levi-Civita connection associated with the metric $\vec{g}$.
\begin{rem}{\em
The choice of the
connection is not arbitrary. As is well-known from  Riemannian geometry, given a connection $\nabla$ on a manifold $M$ one
can parallel propagate frames. For any path $\tau$ between two points of $M$ parallel transport along $\tau$
defines a linear  mapping $L(\tau)$ between the tangent spaces of two points. This linear map is an {\em isometry} if
the connection $\nabla$ is a Levi-Civita connectiction. 
Clearly, the linear map $L(\tau)$ induced by a Levi-Civita connection $\nabla$
maps orthonormal frames to orthonormal frames. }
\end{rem}
The vanishing of
the torsion tensor $T^a{}_{bc}$ is given by  
\begin{equation}
T^a{}_{bc} = \Gamma_{bc}{}^a - \Gamma_{cb}{}^a - C^a{}_{bc} =0,
\label{T}
\end{equation}
while the componets of  the  Riemann curvature tensor
$R^a{}_{bcd}$ with respect to to the moving frame are defined as follows:
\begin{equation}
R^a{}_{bcd} =  E_c\Gamma_{db}{}^a + \Gamma_{db}{}^e\Gamma_{ce}{}^a - E_d\Gamma_{cb}{}^a 
 - \Gamma_{cb}{}^e\Gamma_{de}^a - C^e{}_{cd}\Gamma_{eb}{}^a,
\label{R}
\end{equation}
respectively. We now define a one-form valued   matrix  $\omega^a{}_b$ called the {\it connection one-form} by
\begin{equation}
\label{c4}
\omega^a{}_b : = \Gamma_{cb}{}^aE^c.
\end{equation}
Further, we can define 
$$\omega_{ab}:= g_{ad}\omega^d{}_b.$$
On account of the above the connection one-forms, $\omega_{ab}$ are obviously
skew-symmetric. They satisfy Cartan's structural equations,
\begin{equation}
\label{cs1}
dE^a + \omega^a{}_b\wedge E^b = 0,
\end{equation}
\begin{equation}
\label{cs2}
d\omega^a{}_b + \omega^a{}_c\wedge \omega^c{}_b = \Theta^a{}_b,
\end{equation}
where we have introduced the
{\em curvature two-form} $\Theta^a{}_b := (1/2) R^a{}_{bcd}E^c\wedge E^d.$
Taking the exterior derivative of $(\ref{cs1})$ and $(\ref{cs2})$ yields 
the first
\begin{equation}
\label{*cb1}
 \Theta^a{}_b\wedge E^b = 0
\end{equation}
and second
\begin{equation}
\label{*cb2}
d\Theta^a{}_b + \omega^a{}_c\wedge \Theta^c{}_b - \Theta^a{}_c\wedge \omega^c{}_b = 0
\end{equation}
Bianchi identities, respectively. In addition, for a $(0,2)$ Killing tensor $\vec{K}$ we have the Killing tensor equation:
\begin{equation}
\label{K0}
K_{(ab;c)} = 0,
\end{equation}
where $;$ denotes the covariant derivative defined by
\begin{equation}
\label{K1}
K_{ab;c} := E_cK_{ab} - K_{db}\Gamma_{ca}{}^d - K_{ad}
\Gamma_{cb}{}^d.
\end{equation}
Note that the equation (\ref{K0}) is  the covariant version of the Killing tensor equation given in
terms of the Schouten tensor for contravairant Killing tensors (\ref{KT}). 
We now adapt the equations of Cartan's geometry listed above to the study of the vector space ${\cal K}^2(\mathbb{E}^2)$.
Define $\Gamma_{abc}: = g_{cd}\Gamma_{ab}{}^d$, $C_{abc}:=g_{ad}C^{d}{}_{bc}$ and $R_{abcd}:= g_{ae}R^{e}{}_{bcd}$. In  the case of a two-dimensional (pseudo)-Riemannian
manifold $(M,\vec{g})$ there are only two linearly independent connection coefficients $\Gamma_{112}$,  $\Gamma_{212}$
and one component of the Riemann tensor $R_{1212}$.
For convenience  we write $\gamma:= \Gamma_{112}$ and  $\delta:= \Gamma_{212}$. 
Since in our case $(M,\vec{g}) = \mathbb{E}^2$, 
the condition of flatness $R_{1212} = 0$ applied to (\ref{R}) in  the new notations yields: 
\begin{equation}
\label{RF}
R_{1212} = -E_1\delta + E_2\gamma - \gamma^2 - \delta^2 = 0.
\end{equation}
The equations (\ref{C}) and (\ref{cs1}) transform accordingly:
\begin{equation}
\label{CMF}
[E_1,E_2] = - \gamma E_1 - \delta E_2,
\end{equation}
\begin{equation}
\label{semf}
dE^1 = \gamma E^1\wedge E^2, \quad dE^2 = \delta E^1 \wedge E^2.
\end{equation}
In a similar way, the Killing tensor equation (\ref{K0}) yields the following equations:
\begin{equation}
\label{KTmf}
E_2\lambda_1 = 2\gamma (\lambda_1 - \lambda_1),\quad E_1\lambda_1 = 2\delta (\lambda_2-\lambda_1),\quad E_1\lambda_1 = E_2\lambda_2 = 0,
\end{equation}
where (\ref{T}) has been used. The conditions for orthogonal integrability for $E_1$ and $E_2$ follow from (\ref{semf}): 
\begin{equation}
\label{oi}
E^a\wedge dE^a = 0, \quad a = 1,2.
\end{equation}
Therefore by  Frobenius'  theorem there exist functions $f$, $g$ and variables $u$, $v$, such that
\begin{equation}
\label{funct}
E^1 = fdu, \quad E^1 = gdv.
\end{equation}
In addition, applying $[E_1,E_2]$ to the eigenvalues $\lambda_1$ and $\lambda_2$ we arrive at the integrability
conditions
\begin{equation}
\label{ic}
E_1\gamma = -3\gamma\delta, \quad E_2\delta = 3\gamma\delta.
\end{equation}
It is easy to show (for more details see \cite{BMS01}) that 
\begin{equation}
\gamma = -\frac{1}{fg}\frac{\partial f}{\partial u}, \quad \delta = \frac{1}{fg}\frac{\partial g}{\partial v}.
\label{gd}
\end{equation}
Moreover, $f = g$ and 
\begin{equation}
\label{f}
f^2(u,v) = A(u) + B(v).
\end{equation} 
Now to  obtain the expressions for the canonical  forms of the four orbits it is natural to consider the
following three {\em isometrically invariant} cases

\begin{equation}
\label{cases}
\begin{array}{lrcl}
\mbox{Case 1} & \gamma = \delta = 0 & \Leftrightarrow & \lambda_1 \, \mbox{and}\, \lambda_2 \, \mbox{are constant}\\[0.3cm]
\mbox{Case 2} & \gamma=0, \delta \not=0 \, (\gamma\not=0, \delta = 0) & \Leftrightarrow & \lambda_1\, \mbox{is constant}\,(\lambda_2\,
\mbox{is constant}) \\ [0.3cm]
\mbox{Case 3} & \gamma\delta \not = 0 & \Leftrightarrow & \lambda_1 \, \mbox{and}\, \lambda_2 \, \mbox{are not constant}
\end{array}
\end{equation}

The first step is to solve equations (\ref{KTmf}) in terms of the variables $u$ and $v$, which reduces to
finding the corresponding expressions for the eigenvalues $\lambda_1$ and $\lambda_2$. Upon integration, substituting the formulas (\ref{gd}) and (\ref{f}) into (\ref{KTmf})  and then transforming the result to its contravariant form, we obtain the following fundamental formula (for more
details see \cite{BMS01}): 
\begin{equation}
\label{KTcf}
\vec{K} = \ell_1\vec{K}_c + \ell_2\vec{g}, \quad \ell_1, \ell_2 \in \mathbb{R},
\end{equation}
where 
\begin{equation}
\label{Kc}
\vec{K}_c = \frac{1}{A(u) + B(v)}\left(B(v) \partial_u \odot \partial_u  
-A(u)\partial_v \odot \partial_v\right)
\end{equation}
and
\begin{equation}
\label{Mc}
\vec{g} = \frac{1}{A(u) + B(v)}\left(\partial_u\odot \partial_u + \partial_v\odot \partial_v\right)
\end{equation}
for some smooth functions $A(u)$ and $B(v)$, which are the same as in (\ref{lm}). The formula (\ref{KTcf}) is the solution
to the Killing tensor equation in the rigid moving frames of eigenvectors (eigenvalues) of the unknown $\vec{K}$. Note it does not depend on the curvature 
of the underlying space. Importantly, (\ref{KTcf}) represents the {\em canonical forms} of the orbits of the action ${\cal K}^2(M)/I(M)$, 
where $(M,\vec{g})$ is a two-dimensional Riemannian manifold of constant curvature, ${\cal K}^2(M)$ is the vector space of Killing two-tensors defined on $(M,\vec{g})$ and $I(M)$ is the corresponding isometry group. 
 To specify the formula (\ref{KTcf}) to the Euclidean plane $\mathbb{E}^2$, we employ the flatness condition (\ref{RF}), and then find the corresponding $A(u)$ and $B(v)$ in each of the three cases (\ref{cases}). The problem reduces to solving 
the corresponding differential equations determined by (\ref{RF}). Each of Case 1 and 2 leads to one solution and Case 3 yields two distinct solutions. Thus, as expected, in total 
we arrive at four formulas for $\vec{K}_c$ corresponding to four differnt types of orbits of the orbit space ${\cal K}^2(\mathbb{E}^2)/I(\mathbb{E}^2)$ \cite{BMS01}:
\begin{equation}
\label{cKT}
\begin{array}{rl}
\mbox{Cartesian (C)}: & \displaystyle \vec{K}^C_c = \partial_u \odot \partial_u, \\ [0.3cm]
\mbox{Polar (P)}: &  \vec{K}^P_c =  \partial_v \odot \partial_v, \\[0.3cm]
\mbox{Parabolic (PB)}: & \displaystyle \vec{K}^{PB}_c = \frac{v^2\partial_u\odot \partial_u - u^2\partial_v\odot \partial_v}{u^2+v^2}, \\[0.3cm]
\mbox{Elliptic-hyperbolic (EH)}: & \displaystyle  \vec{K}^{EH}_c = \frac{\cos^2v\partial_u\odot\partial_u +\cosh^2u\partial_v\odot\partial_v}{k^2(\cosh^2u - \cos^2v)}. 
\end{array}
\end{equation}
Accordingly, the metric (\ref{Mc}) that appears in the general formula for the canonical forms (\ref{KTcf}), in each of the
four cases is given by
\begin{equation}
\label{cM}
\begin{array}{rl}
\mbox{Cartesian (C)}: & \displaystyle \vec{g}^C = \partial_u \odot \partial_u + \partial_v \odot \partial_v, \\ [0.3cm]
\mbox{Polar (P)}: & \displaystyle \vec{g}^P = \partial_u\odot \partial_u + \frac{1}{u^2} \partial_v \odot \partial_v, \\[0.3cm]
\mbox{Parabolic (PB)}: & \displaystyle \vec{g}^{PB} = \frac{\partial_u\odot \partial_u + \partial_v\odot \partial_v}{u^2+v^2},\\[0.3cm]
\mbox{Elliptic-hyperbolic (EH)}: & \displaystyle  \vec{g}^{EH} = \frac{\partial_u\odot\partial_u +\partial_v\odot\partial_v}{k^2(\cosh^2u - \cos^2v)}. 
\end{array}
\end{equation}
We next compare the metric (\ref{EM}) given in terms of the Cartesian coordinates $(q_1,q_2)$ with the corresponding four metrics given by (\ref{cM}) in terms of the coordinates $(u,v)$ and employ the method introduced recently in \cite{HM06} to derive in each case the transformations from the coordinates $(u,v)$ to the Cartesian coordinates $(q_1,q_2)$: 
\begin{equation}
\label{TR}
\begin{array}{rl}
\mbox{Cartesian (C)}: & q_1 = u, \quad q_2 = v, \\ [0.3cm]
\mbox{Polar (P)}: & q_1 = u\cos v, \quad  q_2 = u\sin v,\\[0.3cm]
\mbox{Parabolic (PB)}: & q_1 = 1/2(u^2 - v^2), \quad  q_2 = uv,  \\[0.3cm] 
\mbox{Elliptic-hyperbolic (EH)}: & q_1 = k\cosh u \cos v, \, q_2 = k \sinh u \sin v. 
\end{array}
\end{equation}
Finally, we use the formulas (\ref{TR}) to transform back to the Cartesian coordiantes $(q_1,q_2)$ 
  the canonical Killing tensors given by (\ref{cKT}): 
 \begin{equation}
\label{CKT}
\begin{array}{rccl}
\mbox{Cartesian (C)}: & \displaystyle \vec{K}^C_c &  = & \partial_1 \odot \partial_1, \\ [0.3cm]
\mbox{Polar (P)}: &  \vec{K}^P_c & = & q_2^2\partial_1\odot\partial_1 -q_1q_2\partial_1\odot\partial_2 + q_1^1\partial_2\odot\partial_2, \\[0.3cm]
\mbox{Parabolic (PB)}: & \vec{K}^{PB}_c & = & q_2\partial_1\odot\partial_2 - 2q_1\partial_2\odot \partial_2, \\[0.3cm]
\mbox{Elliptic-hyperbolic (EH)}: &  \vec{K}^{EH}_c & = & (k^2 + q_2^2)\partial_1\odot \partial_1 - q_1q_2\partial_1\odot\partial_2 \\ [0.3cm]
& & & + q_1^2\partial_2\odot \partial_2.
\end{array}
\end{equation} 
Now, we can describe the canonical forms of the orbits of every type. For example, a general formula of the canonical forms of the orbits whose elements generate the elliptic-hyperbolic systems of coordiantes can be obtained by substituting into
the formula (\ref{KTcf}) the cooresponding expression for the canonical Killing tensor $\vec{K}_c^{EH}$ given by 
(\ref{CKT}) and the formula  (\ref{EM}) for $\vec{g}$. Therefore, in terms of the Cartesian coordinates $(q_1, q_2)$ the 
canonical forms of the elliptic-hyperbolic orbits are given by the following family of Killing tensors: 
\begin{equation}
\label{CEH}
\vec{K}^{EH} = \ell_1\vec{K}_c^{EH} + \ell_2\vec{g}, \quad \ell_1,\ell_2\in \mathbb{R},
\end{equation}
 where $\vec{K}_c^{EH}$ and $\vec{g}$ are given by (\ref{CKT}) and (\ref{EM}) respectively. We immediately see that, for example, 
  the Killing tensor
 (\ref{ccK}) belongs to the family (\ref{CEH}) for $\ell_1 = 1$, $\ell_2 = 0$, $\beta_1 - \beta_2 = k^2$ and as such represents
 a canonical form of the corresponding elliptic-hyperbolic orbit (see also \cite{JM05}). 
 \begin{rem} {\rm 
 We note that as far as applications of these results to the study of Hamiltonian systems are concerned, it  suffices to 
 study only the  $\vec{K}_c$ part of the formula (\ref{KTcf}). Indeed, if the Hamiltonian system defined by the Hamiltonian $H$ (\ref{H2})
 admits a first integral $F$ (\ref{fi}), then it also admits a first integral of the form $\ell_1H + \ell_2F,$ $\ell_1,\ell_2 \in 
 \mathbb{R}$ and vice versa. }
 \end{rem} 
 We have solved the canonical forms  problem for the orbit space ${\cal K}^2(\mathbb{E}^2)/
 I(\mathbb{E}^2)$. More specifically, for each type of orbits we have derived the most general form of the canonical forms. However, 
 as we have seen in Example \ref{exa1}, in applications, such as the Betrand-Darboux problem, non-trivial Killing tensors can appear not
 in their canonical forms (unless, of course, $V = 0$ in (\ref{H2})). Thus, one has to be able to solve the equivalence forms problem, namely for a given non-trivial Killing tensor (which may not be in its canonical form!), find a way to determine {\em invariantly}
 what type of orbit it belongs to. When it is done, one has to develop a systematic way of transforming a given non-trivial Killing
 tensor to its respective canonical form, as specified by the formula (\ref{KTcf}). To achieve these goals, we shall employ another
 version of the moving frames method that has been introduced and developed recently \cite{FO98, FO99, PO99}. In our case,  the idea is to identify
 the bundle of frames of eigenvectors of non-trivial Killing two-tensors with the group action $I(\mathbb{E}^2) \circlearrowright {\cal K}^2(\mathbb{E}^2)$ (since the group acts transitively on the bundle of frames) and then work in the group. We begin by determining
 the action of the Lie group $I(\mathbb{E}^2)$ in the vector space ${\cal K}^2(\mathbb{E}^2)$ given by a general element (\ref{gKt}). This action is induced by the action  $I(\mathbb{E}^2) \circlearrowright \mathbb{E}^2$ given by
 \begin{equation}
\begin{array}{l}
\tilde{q}_1 = q_1\cos p_3 - q_2\sin p_3 + p_1, \\ [0.3cm]
\tilde{q}_2 = q_1\sin p_3 + q_2\cos p_3 + p_2, 
\end{array}
\label{A1}
\end{equation}
where $p_1,p_2$ and $p_3$ are the parameters of the isometry group $I(\mathbb{E}^2)$. Using (\ref{A1}) and the standard tensor transformation laws applied to the general Killing tensor (\ref{gKt}), we arrive at the following formulas:
 \begin{equation}
\label{A2}
\begin{array}{rcl}
\tilde{\beta_1} &=& \beta_1\cos^2p_3 - 2\beta_3\cos p_3\sin p_3 + \beta_2\sin^2p_3 - 2p_2\beta_4\cos p_3 - 2p_2\beta_5\sin p_3
\\
                  & & + \beta_6p_2^2, \\ 
\tilde{\beta_2} & =& \beta_1\sin^2p_3 - 2\beta_3\cos p_3\sin p_3 +
\beta_2\cos^2p_3 - 2p_1\beta_5\cos p_3 + 2p_1\beta_4\sin p_3 \\ 
 & & + \beta_6p_1^2,\\
\tilde{\beta_3} & = & (\beta_1-\beta_2)\sin p_3\cos p_3 + \beta_3(\cos^2p_3 - \sin^2p_3) 
+ (p_1\beta_4 + p_2\beta_5)\cos p_3 \\ 
 & &  + (p_1\beta_5 - p_2\beta_4)\sin p_3 - \beta_6p_1p_2, \\
 \tilde{\beta_4} & = & \beta_4\cos p_3 + \beta_5\sin p_3 - \beta_6p_2, \\
 \tilde{\beta_5} & = & \beta_5\cos p_3 - \beta_4\sin p_3 - \beta_6p_1,\\
 \tilde{\beta_6} &=& \beta_6. 		  
\end{array}
\end{equation} 
The formulas (\ref{A2}) were first derived in \cite{WF65} and then independently rediscovered in \cite{JMP02}. Next, using the standard
techniques of the Lie group theory, we derive the infinitesimal action of the group $I(\mathbb{E}^2)$ in the vector space ${\cal K}^2(\mathbb{E}^2)$ given by the following generators of its Lie algebra \cite{JMP02}: 
\begin{equation}
\label{gen2}
\begin{array}{l}
\vec{V}_1 = \displaystyle -2\beta_5\frac{\partial}{\partial \beta_2} - \beta_4\frac{\partial}{\partial \beta_3}
 + \beta_6\frac{\partial}{\partial \beta_5},
\\ [0.3cm]
\vec{V}_2 = \displaystyle 2\beta_4\frac{\partial}{\partial \beta_1} - \beta_5\frac{\partial}{\partial \beta_3} 
+ \beta_6\frac{\partial}{\partial \beta_4}, 
\\ [0.3cm]
\vec{V}_3 = \displaystyle  -2\beta_3\Big(\frac{\partial}{\partial \beta_1} - \frac{\partial}{\partial \beta_2}\Big) + (\beta_1 - \beta_2)\frac{\partial}{\partial \beta_3}
+ \beta_5\frac{\partial}{\partial \beta_4} - \beta_4\frac{\partial}{\partial \beta_5}.
\end{array}
\end{equation}
It follows from the formulas (\ref{gen2}) that the dimension of the orbits, which is determined at  each point by the
number of linearly independent vector fields (\ref{gen2}), varies from $0$ to $3$. For example, when $\beta_1 = \beta_2$ and
$\beta_3 = \beta_4 = \beta_5 = \beta_6 = 0$ the dimension of the orbits is obviously $0$. 

To solve the equivalence problem completely, we follow the approach introduced in \cite{DHMS} (see also \cite{JM05}). First, we choose a
cross-section $K$ that intersects $3$-dimensional orbits transversally where the group action is free and regular: 
\begin{equation}
\label{CS}
K = \{\beta_3 = \beta_4 = \beta_5 = 0\}. 
\end{equation}
Next, we observe that chosing a cross-section in this setting is equivalent to chosing a rigid moving frame in the previous 
considerations, where we employed a classical version of the moving frames method. Moreover, the
intersection points are precisely the canonical forms whose coordinates are given by {\em invariants} of the group action
$I(\mathbb{E}^2) \circlearrowright {\cal K}^2(\mathbb{E}^2)$, that is the functions of the parameters
 $\beta_1, \ldots, \beta_6$ that remain unchanged under the group action (see \cite{JMP02} for more details).
 Next, we derive the {\em moving frame map} $\gamma: {\cal K}^2(\mathbb{E}^2) \rightarrow I(\mathbb{E}^2)$  for
 the {\em normalization equations} corresponding to the cross-section (\ref{CS}):
\begin{equation}
\label{NE}
\tilde{\beta}_3 = \tilde{\beta}_4 = \tilde{\beta}_5 = 0.
\end{equation}
 Indeed, solving (\ref{NE}) for the group parameters $p_1, p_2$ and $p_3$, we get \cite{DHMS} 
\begin{equation}
\label{MFM} 
\begin{array}{rcl}
p_1 & = &\displaystyle  \frac{\beta_5\cos p_3 - \beta_4\sin p_3}{\beta_6}, \\ [0.3cm] 
p_2 & = & \displaystyle \frac{\beta_4\cos p_3 + \beta_5\sin p_3}{\beta_6}, \\ [0.3cm]
p_3 & = & \displaystyle \frac{1}{2}\arctan \frac{2(\beta_3\beta_6 + \beta_4\beta_5)}{\beta_6(\beta_1-\beta_2) - \beta_4^2 + \beta_5^2}.
\end{array}
\end{equation}
Note, the moving frame map $\gamma: {\cal K}^2(\mathbb{E}^2) \rightarrow I(\mathbb{E}^2)$ given by (\ref{MFM}) maps elements of the corresponding orbits to their respective canonical forms. We first observe from (\ref{A2}) that 
\begin{equation}
\Delta_1 = \beta_6 \label{I1}
\end{equation}
is an $I(\mathbb{E}^2)$-invariant of the vector space ${\cal K}^2(\mathbb{E}^2)$. Substituting the expressions for $p_1$, $p_2$ and $p_3$ into the first two formulas in (\ref{A2}), we arrive at two additional {\em fundamental $I(\mathbb{E}^2)$-invariants}: 
\begin{equation}
\label{I23}
\begin{array}{rcl}
\Delta_2 & = & \beta_6(\beta_1 + \beta_2) - \beta_4^2 - \beta_5^2, \\ [0.3cm] 
\Delta_3 & = & (\beta_6(\beta_1 - \beta_2) - \beta_4^2 + \beta_5^2)^2 + 4(\beta_6 \beta_3 + \beta_4 \beta_5)^2.
\end{array}
\end{equation}
Recall that the fundamental invariants $\Delta_1$ and $\Delta_3$ were derived first in \cite{WF65}, and then rediscovered more
systematically, using the method of infinitesimal generators, in \cite{JMP02} (see also \cite{CDM05} for other methods). The fundamental invariants $\Delta_1$ and $\Delta_3$ can 
be used to distinquish between the orbits \cite{WF65, JMP02, JM05}. The classification is as follows.  
\begin{equation}
\label{CL}
\begin{array}{rl}
\mbox{Cartesian (C)}: & \Delta_1 = 0 \quad \Delta_3 = 0,  \\ [0.3cm]
\mbox{Polar (P)}: & \Delta_1 \not= 0 \quad \Delta_3 = 0,   \\[0.3cm]
\mbox{Parabolic (PB)}: & \Delta_1 = 0 \quad \Delta_3 \not=0,  \\[0.3cm] 
\mbox{Elliptic-hyperbolic (EH)}: &  \Delta_1 \not= 0, \quad \Delta_3 \not= 0. 
\end{array}
\end{equation}
The fundamental invariants $\Delta_1$ and $\Delta_3$  in the case when both of them are not zero, have an interesting
geometric meaning. The distance $k$ between the focii  of an elliptic-hyperbolic system of coordinates generated by a non-trivial 
Killing tensor, is clearly an $I(\mathbb{E}^2)$-invariant. Hence $k$ is a function of the fundamenal invariants found above. 
The corresponding formula was found in \cite{JMP02} to be
\begin{equation}
\label{focii}
k = \frac{\sqrt{\Delta_3}}{\Delta_1^2}.
\end{equation}
Given a non-trivial Killing tensor $\vec{K} = {\cal K}^2 (\mathbb{E}^2)$, we compare it with the general form (\ref{gKt}), obtain the values for the corresponding parameters $\beta_1, \ldots, \beta_6$ and then compute the funamental $I(\mathbb{E}^2)$-invariants $\Delta_1$ and $\Delta_3$. The problem of the determination of the orbit type for $\vec{K}$, using the classification (\ref{CL}),  is 
then straightforward. If $\vec{K}$ generate an elliptic-hyperbolic coordinate system, then we compute the corresponding
moving frame map (\ref{MFM}), find the group parameters $p_1$, $p_2$, $p_3$,  and then subsitutte them into the formulas for the 
group action (\ref{A1}). The resulting transformation maps $\vec{K}$ to its canonical form. 

Dealing with the remaining three cases, namely Cartesian, polar and parabolic, is a delicate matter. The group action $I(\mathbb{E}^2) \circlearrowright {\cal K}^2(\mathbb{E}^2)$ is not regular 
globally. Thus, for example,  the vanishing $\Delta_3 = 0$ indicates that the dimension of the orbits drops from three to two. If, in addition, $\Delta_1 = 0$ the dimension of the orbits drops further to one. On the other hand, comparing the expressions for the generators of the Lie algebra of the isometry group $I(\mathbb{E}^2)$ (\ref{gen2}) with the formulas for the canonical forms (\ref{CKT}), we conclude that the Cartesian orbits are $1$-dimensional, polar - $2$-dimensional and the orbits of the  parabolic and elliptic-hyperbolic types are $3$-dimensional. Clearly, in order to derive  moving frame maps for the $1$-dimensional and $2$-dimensional and $3$-dimensional orbits, other than the elliptic-hyperbolic type, one  has to choose different cross-sections (for more details, see \cite{DT03, JM05}). 

Finally, we will solve the same problem considered in Example \ref{exa1}, using the results obtained in this section. 
\begin{exa} [2nd Integrable Case of Yatsun] \label{exa2}
{\rm  This time we will  not find the solution by solving the Betrand-Darboux PDE (\ref{BDE})  for
the  potential $V$.  Instead we will deal with the Killing tensor (\ref{KY}) only. 
The material presented in Section \ref{CG} suggests that once the Killing tensor of (\ref{fi}) is available, solving the Bertrand-Darboux PDE (\ref{BDE})
for $V$ is {\em redundant}. Indeed, now we compare the Killing tensor (\ref{KY}) with the general form (\ref{gKt}) 
to obtain
$$\beta_1 = \frac{3}{4}, \quad \beta_2 = 0, \quad \beta_3 = 0, \quad \beta_4 = -\frac{1}{2},\quad  \beta_5 = 0, \quad \beta_6 = 1. $$
Substituting this data into the formulas for $\Delta_1$ (\ref{I1}) and $\Delta_3$ (\ref{I23}), we obtain
$$\Delta_1 = 1 \not= 0, \quad \Delta_3 = \frac{1}{4} \not=0,$$
which immediately shows that the Killing tensor (\ref{KY}) generates the elliptic-hyperbolic coordinates. Moreover, by (\ref{focii})
the distance between the focii $k = 1$. 
 Next, we compute the
moving frames map (\ref{MFM}): 
$$p_1 = - \frac{1}{2}, \quad p_2 = p_3 = 0.$$
Taking into account the formulas (\ref{A1}) ($q_1 = \tilde{q}_1 + \frac{1}{2}$)  and (\ref{TR}) (for (EH)),
 we conclude that the transformation to separable coordinates is given by (\ref{ct}) as expected. 
Thus, we have solved the same problem  as in Example \ref{exa1} {\em without solving the Bertrand-Darboux PDE
 (\ref{BDE})!} Moreover, the procedure based on the moving frame method is purely
 algorithmical and so it can be   implemented in a
 computer algebra package \cite{H05}. }
\end{exa}

\section{Conclusions}

The analysis presented above (compare Examples \ref{exa1} and \ref{exa2}) demonstrates that the Bertrand-Darboux
problem (and its generalizations \cite{CMP05}) can be solved within the framework of Cartan's geometry, in particular -
its most important asset -  the moving frames 
method. Moreover, the approach, based on the method of moving frames, 
 is algorithmical, independent of the curvature or signature of the underlying pseudo-Riemannian
manifold $(M,\vec{g})$, easily adaptable to different geometric settings. 

\bigskip 

{\bf Acknowledgements}. The author acknowledges with gratitude that he has learned the moving frames method in its
classical and modern formulations from Ray McLenaghan and Peter Olver respectively. The work was supported in 
part by an NSERC Discovery Grant.

\end{document}